\documentclass[10pt,cite,epsf,epsfig]{article}
\usepackage{epsfig}
\usepackage{float}
\usepackage{cite}
\usepackage{amsmath}
\usepackage{amssymb}
\usepackage{epsfig,subfigure}
\usepackage{graphicx}
\usepackage{bbm}
\usepackage{enumerate}
\usepackage{wasysym}

\begin{document}
\author{S. Dev \thanks{sdev@associates.iucaa.in} $^{,1}$,
Desh Raj \thanks{raj.physics88@gmail.com} $^{,2}$}
\date{$^1$\textit{Department of Physics, School of Sciences, HNBG Central University, Srinagar, Uttarakhand 246174, INDIA.}\\
\smallskip
$^2$\textit{Department of Physics, Himachal Pradesh University, Shimla 171005, INDIA.}}
\title{Phenomenology of Trimaximal mixing with one texture equality}
\maketitle
\begin{abstract}
We study neutrino mass matrices with one texture equality and the neutrino mixing matrix having either its first (TM$_1$) or second (TM$_2$) column identical to that of the tribimaximal mixing matrix. We found that out of total fifteen possible neutrino mass matrices with one texture equality, only six textures are compatible with $TM_1$ mixing and six textures are compatible with $TM_2$ mixing in the light of the current neutrino oscillation data. These textures have interesting implications for the presently unknown parameters such as the neutrino mass scale, effective Majorana neutrino mass, effective neutrino mass, the atmospheric mixing and the Dirac- and Majorana-type CP violating phases. We, also, present the $S_{3}$ group motivation for some of these textures.   
\end{abstract}
\section{Introduction}
In the last two decades, significant advances have been made by various neutrino oscillation experiments in determining the neutrino masses and mixings. Various neutrino parameters like three mixing (solar, atmospheric and reactor) angles and the two mass squared differences ($\Delta m^{2}_{21}$ and $|\Delta m^{2}_{31}|$) have been measured by various neutrino oscillation experiments with fairly good precision. In addition, the recent neutrino oscillation data hint towards a nonmaximal atmospheric mixing angle ($\theta_{23}$) \cite{t23} and Dirac-type CP-violating phase ($\delta$) near $270^\circ$ \cite{nova}. However, many other attributes like leptonic CP-violation, neutrino mass ordering (normal mass ordering (NO) or inverted mass ordering (IO)), nature of neutrinos (Dirac or Majorana) and absolute neutrino mass scale are still unknown. Furthermore, the origin of lepton flavor structure still remains an open issue. The neutrino mass matrix which encodes the neutrino properties contains several unknown physical parameters. The phenomenological approaches based on Abelian or non-Abelian flavor symmetries can play a significant role in determining the specific texture structure of the neutrino mass matrix with reduced number of independent parameters. Several predictive models such as texture zeros \cite{texture}, vanishing cofactors \cite{minor}, equalities among elements/cofactors \cite{equal} and hybrid textures \cite{hybrid} amongst others can explain the presently available neutrino oscillation data. Since, the presence of texture equalities, just like texture zeros or vanishing cofactors, reduces the number of free parameters in the neutrino mass matrix and, hence, must have a similar predictability as that of texture zeros or vanishing cofactors. In the flavor basis, neutrino mass matrices with one texture equality and two texture equalities have been studied in the literature \cite{equal}. The hybrid textures which combine a texture equality with a texture zero or a vanishing cofactor  have been studied in the literature \cite{hybrid}.\\
In addition, discrete non-Abelian symmetries leading to the Tri-Bi-Maximal (TBM) \cite{tbm} neutrino mixing pattern have been widely studied in the literature. The TBM mixing matrix given by
\begin{equation}
U_{TBM}=\left(
\begin{array}{ccc}
 \sqrt{\frac{2}{3}} & \frac{1}{\sqrt{3}} & 0 \\
 -\frac{1}{\sqrt{6}} & \frac{1}{\sqrt{3}} & -\frac{1}{\sqrt{2}} \\
 -\frac{1}{\sqrt{6}} & \frac{1}{\sqrt{3}} & \frac{1}{\sqrt{2}} \\
\end{array}
\right)
\end{equation}         
predicts a vanishing reactor mixing angle $(\theta_{13}\!=\!0)$, maximal atmospheric mixing angle $(\theta_{23}\!=\!\pi/4)$ and the solar mixing angle is predicted to be $(\theta_{12})$ is $\sin^{-1}(1/\sqrt{3})$. However, the nonzero value of $\theta_{13}$ confirmed by various neutrino oscillation experiments underlines the need for necessary modifications to the TBM mixing pattern to make it compatible with the present experimental data \cite{tbm1}. One of the simplest possibilities is to keep one of the columns of the TBM mixing matrix unchanged while modifying its remaining two columns to within the unitarity constraints. This gives rise to three mixing patterns viz. $TM_{1}$, $TM_{2}$ and $TM_{3}$ which have their first, second and third columns identical to the TBM mixing matrix, respectively.
The $TM_{3}$ mixing scheme predicts $\theta_{13}=0$ and is, hence, phenomenologically unviable. The $TM_{1}$ and $TM_{2}$ mixing schemes have been, successfully, employed to explain the pattern of lepton mixing and have been, extensively, studied in the literature \cite{tm,tm1,tm2}. The $TM_{1}$ mixing, in particular, gives a very good fit to the present
neutrino oscillation data. Recently, neutrino mass matrices with texture zero(s) in combination with $TM_{1}$ and $TM_{2}$ mixing have been studied \cite{tmzero}.\\
In the present work, we study a class of neutrino mass matrices having one texture equality with $TM_{1}$ or $TM_{2}$ of the TBM in the neutrino mixing matrix. Neutrino mass matrices having one texture equality along with $TM_{1}$ or $TM_{2}$ of the TBM  have a total of six free parameters and, hence, lead to very predictive textures for the neutrino mass matrices.\\
\begin{table}[h]
\begin{center}
\caption{Fifteen possible texture structures with one equality between two nonzero elements.}
\begin{tabular}{lll}
 \hline \hline
Textures &  & Constraints on elements \\
 \hline
$M_{\nu1}$ & &$M_{ee}$=$M_{e\mu}$\\ 
$M_{\nu2}$ & &$M_{ee}$=$M_{e\tau}$\\
$M_{\nu3}$ & &$M_{e\mu}$=$M_{\mu\mu}$\\
$M_{\nu4}$ & &$M_{\mu\mu}$=$M_{\mu\tau}$\\
$M_{\nu5}$ & &$M_{e\tau}$=$M_{\tau\tau}$\\
$M_{\nu6}$ & &$M_{\mu\tau}$=$M_{\tau\tau}$\\
$M_{\nu7}$ & &$M_{ee}$=$M_{\mu\tau}$\\
$M_{\nu8}$ & &$M_{e\tau}$=$M_{\mu\mu}$\\
$M_{\nu9}$ & &$M_{e\mu}$=$M_{\tau\tau}$\\
$M_{\nu10}$ & &$M_{ee}$=$M_{\mu\mu}$\\
$M_{\nu11}$ & &$M_{ee}$=$M_{\tau\tau}$\\
$M_{\nu12}$ & &$M_{\mu\mu}$=$M_{\tau\tau}$\\
$M_{\nu13}$ & &$M_{e\mu}$=$M_{e\tau}$\\
$M_{\nu14}$ & &$M_{e\mu}$=$M_{\mu\tau}$\\
$M_{\nu15}$ & &$M_{e\tau}$=$M_{\mu\tau}$\\ 
 \hline
 \end{tabular}
\label{table:equal}
\end{center}
\end{table}
 There are a total of fifteen possible structures with one texture equality in the neutrino mass matrix and listed in Table \ref{table:equal}. There exists a $\mu$-$\tau$ permutation symmetry between different structures of neutrino mass matrices and the corresponding permutation matrix has the following form:
\begin{equation}
P_{23}=\left(
\begin{array}{ccc}
 1 & 0 & 0 \\
 0 & 0 & 1 \\
 0 & 1 & 0 \\
\end{array}
\right).
\end{equation}   
Neutrino mass matrices with one texture equality, therefore, are related to each other as
\begin{equation}
M^{'}_{\nu}=P_{23} M_{\nu} P^{T}_{23}
\end{equation}
leading to the following relations between the neutrino oscillation parameters:
\begin{equation}
\theta^{'}_{12}=\theta_{12},~~ \theta^{'}_{13}=\theta_{13},~~ \theta^{'}_{23}=\frac{\pi}{2}-\theta_{23},~~ \delta^{'}=\pi-\delta.
\end{equation}
Neutrino mass matrices with one texture equality related by $\mu$-$\tau$ permutation operation are
\begin{eqnarray}
M_{\nu1}\leftrightarrow M_{\nu2},~ M_{\nu3}\leftrightarrow M_{\nu5}, ~M_{\nu4}\leftrightarrow M_{\nu6},\nonumber\\
M_{\nu7}\leftrightarrow M_{\nu7},~ M_{\nu8}\leftrightarrow M_{\nu9}, ~M_{\nu10}\leftrightarrow M_{\nu11},\\
M_{\nu12}\leftrightarrow M_{\nu12}, ~M_{\nu13}\leftrightarrow M_{\nu13},~ M_{\nu14}\leftrightarrow M_{\nu15}. \nonumber 
\end{eqnarray} 

In the flavor basis, where the charged lepton mass matrix $M_{l}$ is diagonal, the complex symmetric Majorana neutrino mass matrix $M_{\nu}$ can be diagonalized by a unitary matrix $V'$:
\begin{equation}
M_{\nu}=V'M_{\nu}^{diag}V'^{T}
\end{equation}
where $M_{\nu}^{diag}=diag(m_1,m_2,m_3)$. The unitary matrix $V'$ can be parametrized as
\begin{equation}
V'=P_{l} V ~~\textrm{with}~~ V=U P_{\nu}
\end{equation}
where
\begin{equation}
U=\left(
\begin{array}{ccc}
 c_{12} c_{13} & c_{13} s_{12} & e^{-i \delta } s_{13} \\
 -c_{23} s_{12}-e^{i \delta } c_{12} s_{13} s_{23} & c_{12} c_{23}-e^{i \delta } s_{12} s_{13} s_{23}
   & c_{13} s_{23} \\
 s_{12} s_{23}-e^{i \delta } c_{12} c_{23} s_{13} & -e^{i \delta }
  c_{23} s_{12} s_{13}-c_{12} s_{23} & c_{13} c_{23}
\end{array}
\right),\\
\end{equation}
\begin{equation}
P_{\nu}=
\left(
\begin{array}{ccc}
1 & 0 & 0 \\
0 & e^{i \alpha}&0 \\
0 & 0 & e^{i \beta}
\end{array}
\right),~~\textrm{and}~~ P_{l}=
\left(
\begin{array}{ccc}
e^{i \phi_{e}} & 0 & 0 \\
0 & e^{i \phi_{\mu}}&0 \\
0 & 0 & e^{i \phi_{\tau}}
\end{array}
\right),
\end{equation}
with $c_{ij}=\cos \theta_{ij}$, $s_{ij}=\sin \theta_{ij}$. $P_{\nu}$ is the diagonal phase matrix containing the two Majorana-type CP-violating phases $\alpha$ and $\beta$. $\delta$ is the Dirac-type CP-violating phase. The phase matrix $P_l$ is physically unobservable. The matrix $V$ is called the neutrino mixing matrix or the Pontecorvo-Maki-Nakagawa-Sakata (PMNS) \cite{pmns} matrix. The effective Majorana neutrino mass matrix can be written as
\begin{equation}
M_{\nu}=P_{l}U P_{\nu}M_{\nu}^{diag}P_{\nu}^{T}U^{T}P_{l}^{T}.
\end{equation}

The Dirac-type CP-violation in neutrino oscillation experiments can be described in terms of the Jarlskog rephasing invariant quantity $J_{CP}$ \cite{jcp} with
\begin{equation}
J_{CP}=\textrm{Im}\lbrace U_{11} U_{22} U^{*}_{12} U^{*}_{21}\rbrace=\sin\theta_{12} \sin\theta_{23} \sin\theta_{13} \cos\theta_{12} \cos\theta_{23} \cos^{2}\theta_{13} \sin\delta .
\end{equation}
The effective Majorana neutrino mass $|M_{ee}|$, which determines the rate of neutrinoless double beta decay, is given by
\begin{equation}
|M_{ee}|=|m_1 U^{2}_{e1}+m_2 U^{2}_{e2}+m_3 U^{2}_{e3}|.
\end{equation}
There are many experiments such as CUORICINO \cite{cuori}, CUORE \cite{cuore}, MAJORANA \cite{majorana}, SuperNEMO \cite{nemo}, EXO \cite{exo} which aim to achieve a sensitivity upto $0.01$ eV for $|M_{ee}|$. KamLAND-Zen experiment \cite{kam1} provide the upper limits on the effective Majorana neutrino mass which is given by
\begin{equation}
|M_{ee}|< (0.36\textrm{-}0.156) ~\textrm{eV}
\end{equation}
at 90$\%$ confidence level (C.L.).\\
The measurement of the absolute neutrino mass scale via the decay kinematics is usually described by the effective neutrino mass \cite{shro}
\begin{equation}
m_{\beta} \equiv \sqrt{m_1^2 |U_{e1}|^2+m_2^2 |U_{e2}|^2+m_3^2 |U_{e3}|^2}
\end{equation}
Recently, the KATRIN \cite{katrin} experiment has reported the upper limit of $m_{\beta} < 0.8$ eV at 90$\%$ C.L.\\  
Further, cosmological observations provide more stringent constraints on absolute neutrino mass scale by putting an upper bound on the sum of neutrino masses:
\begin{eqnarray}
\sum=\sum\limits_{i=1}^{3} m_{i}.
\end{eqnarray}
Recent Planck data \cite{planck} in combination with baryon acoustic oscillation (BAO) measurements provide a tight bound on the sum of neutrino masses $\sum m_{i}\leq 0.12$ eV at 95$\%$ C.L. 
\begin{table}[h]
\caption{Current Neutrino oscillation parameters from global fits \cite{data} with $\Delta m^{2}_{3 l}\equiv \Delta m^{2}_{31}>0$ for NO and $\Delta m^{2}_{3 l}\equiv \Delta m^{2}_{32}=-\Delta m^2_{23}<0$ for IO.}
\begin{center}
\begin{tabular}{lll}
 \hline \hline
Neutrino Parameter & Normal Ordering (best fit)& Inverted Ordering ($\Delta \chi^2=2.6$)\\
  & bfp $\pm 1 \sigma$ ~~~~~~~~~ $3\sigma$ range  & bfp $\pm 1 \sigma$~~~~~~~~~~ $3\sigma$ range \\
 \hline
$\theta_{12}^{\circ}$ & $33.44^{+0.77}_{-0.74}$ ~~~~~~ $31.27 \rightarrow 35.86$ & $33.45^{+0.77}_{-0.74}$ ~~~~~~~ $31.27 \rightarrow 35.87$ \\
$\theta_{23}^{\circ}$ & $49.2^{+1.0}_{-1.3}$  ~~~~~~~~~~~$39.5 \rightarrow 52.0$ & $49.5^{+1.0}_{-1.2}$ ~~~~~~~~~~$39.8 \rightarrow 52.1$ \\
$\theta_{13}^{\circ}$ & $8.57^{+0.13}_{-0.12}$ ~~~~~ ~~~~$8.20 \rightarrow 8.97$ & $8.60^{+0.12}_{-0.12}$ ~~~~~~~~ $8.24 \rightarrow 8.98$ \\
$\delta_{CP}^{\circ}$ & $194^{+52}_{-25}$ ~~~~~~~~~~~~ $105 \rightarrow 405$ & $287^{+ 27}_{-32}$ ~~~~~~~~~~~ $192 \rightarrow 361$ \\
$\Delta m^{2}_{21}/10^{-5} eV^2 $ & $7.42^{+0.21}_{-0.20}$ ~~~~~~~~~~$6.82 \rightarrow 8.04$ & $7.42^{+0.21}_{-0.20}$ ~~~~~~~~ $6.82 \rightarrow 8.04$ \\
$\Delta m^{2}_{3 l}/10^{-3} eV^2 $ & $+2.515^{+0.028}_{-0.028}$ ~ $
+2.431 \rightarrow +2.599$ & $-2.498^{+0.028}_{-0.029}$~  $-2.584 \rightarrow -2.413$ \\
 \hline 
 \end{tabular}
\label{table:dat}
\end{center}
\end{table}

\section{$TM_{2}$ mixing and one texture equality}

A neutrino mass matrix with $TM_{2}$ mixing can be written as
\begin{equation}
M_{TM_{2}}=P_{l}U_{TM_{2}} P_{\nu}M_{\nu}^{diag}P_{\nu}^{T}U_{TM_{2}}^{T}P_{l}^{T}
\end{equation}
where the mixing matrix $TM_{2}$, also known as trimaximal mixing, can be parametrized \cite{tm2} as 
\begin{equation}
U_{TM_{2}}=\left(
\begin{array}{ccc}
 \sqrt{\frac{2}{3}} \cos\theta & \frac{1}{\sqrt{3}} & \sqrt{\frac{2}{3}} \sin\theta \\
 \frac{e^{i \phi } \sin\theta}{\sqrt{2}}-\frac{\cos\theta}{\sqrt{6}} &
   \frac{1}{\sqrt{3}} & -\frac{e^{i \phi } \cos\theta}{\sqrt{2}}-\frac{\sin\theta}{\sqrt{6}} \\
 -\frac{\cos\theta}{\sqrt{6}}-\frac{e^{i \phi } \sin\theta}{\sqrt{2}} &
   \frac{1}{\sqrt{3}} & \frac{e^{i \phi } \cos\theta}{\sqrt{2}}-\frac{\sin\theta}{\sqrt{6}} \\
\end{array}
\right).
\end{equation}
The mass matrix $M_{TM_{2}}$ is invariant under the transformation $G_{2}^{T}M_{TM_{2}}G_{2}=M_{TM_{2}}$
 with $G_{2}=U_{TM_2}diag(-1,1,-1)U^\dagger_{TM_2}$, as the generator of $Z_{2}$ symmetry \cite{ge}. Invariance of $M_{TM_{2}}$ under $G_{2}$ when combined with one texture equality leads to the equality of three unphysical phases in $M_{TM_{2}}$ i.e., $\phi_{e}=\phi_{\mu}=\phi_{\tau}\equiv\phi_{l}$.\\
The most general neutrino mass matrix with $TM_{2}$ as the mixing matrix can be parametrized as
\begin{eqnarray}
M_{TM_{2}}=\left(
\begin{array}{ccc}
 u+\frac{2 x}{3} & v-\frac{x}{3} & w-\frac{x}{3} \\
 v-\frac{x}{3} & w+\frac{2 x}{3} & u-\frac{x}{3} \\
 w-\frac{x}{3} & u-\frac{x}{3} & v+\frac{2 x}{3} \\
\end{array}
\right)\approx \left(
\begin{array}{ccc}
 a & b & c \\
b & c+d & a-d \\
 c & a-d & b+d \\
\end{array}
\right).
\end{eqnarray} 
The neutrino mass matrix $M_{TM_{2}}$ can be realized within the framework of an $A_{4}$ model where the $A_{4}$ flavor symmetry is spontaneously broken by two real $A_{4}$ triplets $\phi, \phi^\prime$, and three real $A_{4}$ singlets, $\xi,\xi^\prime,\xi^{\prime\prime}$ which are $SU(2)_L$ gauge singlets \cite{honda}. Upon symmetry breaking, the VEVs of the flavon singlets and triplets take the alignments
\begin{eqnarray}
\langle\xi \rangle=u_{a}, \langle\xi^\prime \rangle=u_{c}, \langle\xi^{\prime^\prime} \rangle=u_{b}, \langle\phi\rangle=(v,v,v), \langle\phi^\prime\rangle=(v^{\prime},0,0).
\end{eqnarray}
 The neutrino mass matrix, in the flavor basis, is given by
\begin{eqnarray}
\left(
\begin{array}{ccc}
 u+\frac{2 x}{3} & v-\frac{x}{3} & w-\frac{x}{3} \\
 v-\frac{x}{3} & w+\frac{2 x}{3} & u-\frac{x}{3} \\
 w-\frac{x}{3} & u-\frac{x}{3} & v+\frac{2 x}{3} \\
\end{array}
\right)
\end{eqnarray}
The above mass matrix leads to the $TM_{2}$ neutrino mixing matrix. For $v=w$, the above mass matrix leads to the TBM neutrino mixing matrix. 
The equality among the elements of mass matrix in Eq. (18) does not arise naturally and, hence, we assume additional constraints on the elements of mass matrix e.g. $u=v-x$ which leads to one equality between the (1,1) and (1,2)-elements of $M_{TM_{2}}$ in Eq.(18).  
 Therefore, all possible textures of neutrino mass matrices with $TM_{2}$ mixing and one texture equality are given by
\begin{eqnarray}
M_{TM_{2}}^{1}&=&\left(
\begin{array}{ccc}
 a & a & b \\
 a & b+d & a-d \\
 b & a-d & a+d \\
\end{array}
\right),~~~~~~~~~~~~~~~~~~~
M_{TM_{2}}^{2}=\left(
\begin{array}{ccc}
 a & b & a \\
 b & a+d & a-d \\
 a & a-d & b+d \\
\end{array}
\right),\\
M_{TM_{2}}^{3}&=&\left(
\begin{array}{ccc}
 b+d & a & a-d \\
 a & a & b \\
 a-d & b & a+d \\
\end{array}
\right),~~~~~~~~~~~~~~~~~~~
M_{TM_{2}}^{5}=\left(
\begin{array}{ccc}
 b+d & a-d & a \\
 a-d & a+d & b \\
 a & b & a \\
\end{array}
\right),\\
M_{TM_{2}}^{4}&=&\left(
\begin{array}{ccc}
 a+d & b & a-d \\
 b & a & a \\
 a-d & a & b+d \\
\end{array}
\right),~~~~~~~~~~~~~~~~~~~
M_{TM_{2}}^{6}=\left(
\begin{array}{ccc}
 a+d & a-d & b \\
 a-d & b+d & a \\
 b & a & a \\
\end{array}
\right),\\
M_{TM_{2}}^{7}&=&M_{TM_{2}}^{8}=M_{TM_{2}}^{9}=\left(
\begin{array}{ccc}
 a & b & d \\
 b & d & a \\
 d & a & b \\
\end{array}
\right),\\
M_{TM_{2}}^{10}&=&M_{TM_{2}}^{15}=\left(
\begin{array}{ccc}
 a & b & b-d \\
 b & a & b-d \\
 b-d & b-d & a+d \\
\end{array}
\right),~~~~
M_{TM_{2}}^{11}=M_{TM_{2}}^{14}=\left(
\begin{array}{ccc}
 a & b & b+d \\
 b & a+d & b \\
 b+d & b & a \\
\end{array}
\right),\\
M_{TM_{2}}^{12}&=&M_{TM_{2}}^{13}=\left(
\begin{array}{ccc}
 a & b & b \\
 b & b+d & a-d \\
 b & a-d & b+d \\
\end{array}
\right)
\end{eqnarray}
where the neutrino mass matrices in each equation are related by $\mu$-$\tau$ symmetry. The neutrino mixing angles can be calculated by using the following relations:
\begin{equation}
\sin^{2}\theta_{13}=|U_{13}|^2,~~~ \sin^{2}\theta_{12}=\frac{|U_{12}|^2}{1-|U_{13}|^2}, ~~\textrm{and}~~ \sin^{2}\theta_{23}=\frac{|U_{23}|^2}{1-|U_{13}|^2}. 
\end{equation}
Substituting the elements of $U$ form Eq.(17) into Eq.(27), we get
\begin{eqnarray}
\sin^{2}\theta_{13}=\frac{2}{3}\sin^{2}\theta,~~~ \sin^{2}\theta_{12}=\frac{1}{3-2 \sin^{2}\theta},\\ \nonumber
\textrm{and}~~ \sin^{2}\theta_{23}=\frac{1}{2}\left(1+\frac{\sqrt{3}\sin2\theta \cos\phi}{3-2 \sin^{2}\theta}\right).
\end{eqnarray}
Using Eqs.(11) and (17), the Jarlskog rephasing invariant is given by
\begin{equation}
J_{CP}=\frac{1}{6\sqrt{3}}\sin2\theta \sin\phi,
\end{equation} 
and the Dirac-type CP-violating phase can be calculated by using the equation \cite{tmzero}
\begin{equation}
\tan\delta=\frac{\cos2\theta+2}{2\cos2\theta+1}\tan\phi.
\end{equation}
From Eqs.(12) and (17), the effective Majorana mass for $TM_{2}$ mixing is given by
\begin{equation}
|M_{ee}|=\frac{1}{3}|2m_{1}\cos^{2}\theta + m_{2} e^{2i\alpha}+2m_{3} \sin^{2}\theta e^{2i\beta}|,
\end{equation} 
and the effective neutrino mass for $TM_{2}$ mixing can be calculated by using Eqs.(13) and (17) as
\begin{equation}
m_{\beta} \equiv \sqrt{\frac{1}{3}(2m_{1}^2\cos^{2}\theta + m_{2}^2+2m_{3}^2 \sin^{2}\theta)}
\end{equation}
The existence of one equality between the elements (a,b) and (c,d) of the neutrino mass matrix $M_{TM_{2}}$ implies
\begin{equation}
M_{TM_{2}(ab)}-M_{TM_{2}(cd)}=0
\end{equation}  
which yields the complex equation
\begin{equation}
\sum(Q V_{ai}V_{bi}-V_{ci}V_{di}) ~m_i=0
\end{equation}
where $Q=e^{i(\phi_a+\phi_b-(\phi_c+\phi_d))}$ and $V$ is PMNS matrix given in Eq. (7). The above equation can be rewritten as
\begin{equation}
m_1 A_1+m_2 A_2 e^{2i\alpha}+m_3 A_3 e^{2i\beta}=0
\end{equation}
where
\begin{equation}
A_i=(Q U_{ai}U_{bi}-U_{ci} U_{di})
\end{equation}
with $(i=1,2,3)$ and $a,b,c,d$ can take values $e, \mu$ and $\tau$.
Since the $TM_{2}$ mixing has equal elements in the second column it leads to $A_2\equiv(Q U_{a2}U_{b2}-U_{c2} U_{d2})=0$. Therefore, using Eq. (17) in Eq. (35), we have 
\begin{equation}
m_1 A_1+m_3 A_3 e^{2i\beta}=0.
\end{equation}
Simultaneous solution of the real and imaginary parts of Eq. (37) leads to
\begin{eqnarray}
\xi&\equiv&\frac{m_3}{m_1}=\frac{Re(A_1)}{Im(A_{3})\sin2\beta-Re(A_3)\cos2\beta }=\frac{|A_1|}{|A_3|}, \\
\beta &=&\frac{1}{2}\tan^{-1}\frac{Re(A_3)Im(A_1)-Re(A_1) Im(A_3)}{Re(A_1)Re(A_3)+Im(A_1) Im(A_3)}.
\end{eqnarray}
Using experimentally available mass squared differences $\Delta m^2_{21}$ and $\Delta m^2_{31}~(\Delta m^2_{23})$  for NO (IO) with Eq. (38), the three neutrino mass eigenvalues are given by 
\begin{eqnarray}
&& m_{1}=\sqrt{\frac{\Delta m^2_{31}}{\xi^2-1}}, ~~~~~~~~~~~~m_2=\sqrt{\Delta m^2_{21}+m^2_{1}},~ m_{3}=\xi m_{1}~~~~~~~\textrm{for NO},\\
\textrm{and} ~~&& m_{1}=\sqrt{\frac{\Delta m^2_{23}-\Delta m^2_{21}}{1-\xi^2}},~ m_2=\sqrt{\Delta m^2_{21}+m^2_{1}},~ m_{3}=\xi m_{1} ~~~~~~~\textrm{for IO} 
\end{eqnarray}
where $\Delta m_{ij}^2=m^2_i-m^2_j$, $m_1<m_2<m_3$ for NO and $m_3<m_1<m_2$ for IO.\\
For the numerical analysis, we generate $\sim10^7$-$10^9$ points. The mass squared differences $\Delta m^2_{21}$
and $\Delta m^2_{31}(\Delta m^2_{23})$ for NO (IO) are varied randomly within their 3$\sigma$ experimental ranges given in Table \ref{table:dat}. Parameters $\theta, \phi$,  and $\alpha$ are, also, varied randomly within their full ranges (0-$90^\circ$), (0-$360^\circ$) and (0-$360^\circ$), respectively. Eqs. (39), (40) and (41) are used to calculate the Majorana-type CP-violating phase $\beta$ and three mass eigenvalues ($m_{1}, m_{2}$ and $m_{3}$) for both mass orderings. In addition, the mixing angles $\theta_{12}, \theta_{13}$ and $\theta_{23}$ are calculated by using Eq. (28) and must satisfy the experimental data given in Table \ref{table:dat}. The Jarlskog invariant ($J_{CP}$), Dirac-type CP-violating phase ($\delta$), effective Majorana mass ($|M_{ee}|$), effective neutrino mass ($m_{\beta}$) and the sum of neutrino masses ($\sum$) are calculated by using Eqs. (29), (30), (31), (32) and (15), respectively.\\

The numerical predictions for various neutrino parameters are given in Table \ref{table:tm2} and Table \ref{table:tm3}. Table \ref{table:tm2} provided numerical predictions for viable textures under the constrains from neutrino oscillation data, whereas, Table \ref{table:tm3} provided numerical predictions for viable textures under the constrains from cosmological and neutrinoless double beta decay bounds along with neutrino oscillation data.  The allowed range of parameter $\alpha$ is (0-$360^\circ$) for all viable textures. $J_{CP}$ lies in the range ($-$0.037-0.037) for all viable textures except $M_{TM_{2}}^{3}(M_{TM_{2}}^{5})$ with NO(IO) and for these textures the range of $J_{CP}$ is $\pm(0.011$-$0.037)$. The parameter $\theta$ is constrained to lie within the ranges ($10.0^\circ$-$11.1^\circ$) for all viable textures. The solar mixing angle ($\theta_{12}$) is constrained to lie in the ranges ($35.68^\circ$-$35.77^\circ$) for all allowed textures. Figures 1 and 2 show correlations among various neutrino oscillation parameters. The Dirac CP-violating phase $\delta$ and phase $\phi$ are linearly correlated as shown in Fig. 2(e). Figure 1(h) depicts the correlation between $\theta_{12}$ and $\theta_{13}$. The Dirac-type CP-violating phase $\delta$ strongly depends on the Majorana-type CP-violating phase $\beta$ as shown in Fig. 1(a).

\begin{table}[h]
\begin{center}
\caption{Numerical predictions for viable textures having one equality in $M_{\nu}$ with $TM_{2}$ mixing at $3\sigma$ C.L. (only neutrino oscillation data are incorporated)}
\begin{small}
\begin{tabular}{lllllllll}
\hline\hline
  Texture & Ordering & $m_{\textrm{lowest}}$ (eV)& $|M_{ee}|$ (eV)&$\sum$ (eV)& $m_{\beta}$ & $\theta_{23}^\circ$&$\delta^\circ$&$\beta^\circ$\\
 \hline
$M_{TM_{2}}^{1}$&NO & 0.003-0.0087 & 0.0-0.0083 & 0.062-0.073 & 0.009-0.0128 & 39.5-51.41 & 0-155$\oplus$205-360 &
 16-90$\oplus$270-344\\
$M_{TM_{2}}^{2}$ &NO & 0.003-0.0083 & 0.0-0.0082 & 0.061-0.072 & 0.009-0.0124 & 39.5-51.4 & 0-160$\oplus$202-360 &
 0-81$\oplus$279-360\\
$M_{TM_{2}}^{3}$ & NO & 0.025-0.5 & 0.006-0.5 & 0.10-1.51 &0.026-0.5 & 39.5-45 & 90-160$\oplus$200-270 & 
 0-68$\oplus$292-360\\
&IO & 0.02-0.48 & 0.015-0.5 &  0.12-1.45 & 0.05-0.5 & 45-51.4 & 0-90$\oplus$270-360 & 0-90$\oplus$270-360\\
$M_{TM_{2}}^{4}$ & IO & 0.003-0.008 & 0.014-0.05 & 0.1-0.11 & 0.048-0.051 & 39.8-51.4 & 0-150$\oplus$210-360 &
0-77$\oplus$283-360\\
$M_{TM_{2}}^{5}$ & NO & 0.02-0.5 & 0.005-0.5  & 0.09-1.5 & 0.02-0.5 & 45-51.4 & 0-90$\oplus$270-360 & 0-90$\oplus$270-360\\
& IO & 0.027-0.5 & 0.017-0.47 & 0.13-1.48 & 0.05-0.5 & 39.8-45 & 90-150$\oplus$209-270 &
0-59$\oplus$303-360\\
$M_{TM_{2}}^{6}$ & IO & 0.003-0.009 & 0.014-0.05 & 0.101-0.111 & 0.048-0.051 & 39.8-51.4 & 0-151$\oplus$211-360 & 19-90$\oplus$270-340\\
 \hline
 \end{tabular}
  \end{small}
\label{table:tm2}
\end{center}
\end{table}

\begin{table}[h]
\begin{center}
\caption{Numerical predictions for viable textures having one equality in $M_{\nu}$ with $TM_{2}$ mixing at $3\sigma$ C.L. (cosmological and neutrinoless double beta decay bounds along with neutrino oscillation data are incorporated)}
\begin{small}
\begin{tabular}{lllllllll}
 \hline\hline
Texture & Ordering & $m_{\textrm{lowest}}$ (eV)& $|M_{ee}|$ (eV)&$\sum$ (eV)& $m_{\beta}$ & $\theta_{23}^\circ$&$\delta^\circ$&$\beta^\circ$\\
 \hline
$M_{TM_{2}}^{1}$&NO & 0.003-0.0087 & 0.0-0.0083 & 0.062-0.073 & 0.009-0.0128 & 39.5-51.41 & 0-155$\oplus$205-360 &
 16-90$\oplus$270-344\\
$M_{TM_{2}}^{2}$ &NO & 0.003-0.0083 & 0.0-0.0082 & 0.061-0.072 & 0.009-0.0124 & 39.5-51.4 & 0-160$\oplus$202-360 &
 0-81$\oplus$279-360\\
$M_{TM_{2}}^{3}$ & NO & 0.025-0.031 & 0.006-0.03 & 0.10-0.12 & 0.026-0.032 & 39.5-40.34 & 137-160$\oplus$200-222 & 
 45-68$\oplus$ 292-316\\
$M_{TM_{2}}^{4}$ & IO & 0.003-0.008 & 0.014-0.05 & 0.1-0.111 & 0.048-0.051 & 39.8-51.4 & 0-150$\oplus$210-360 &
0-77$\oplus$283-360\\
$M_{TM_{2}}^{5}$ & NO & 0.021-0.031 & 0.005-0.03  & 0.098-0.12 &0.023-0.032 & 49.7-51.4 & 0-42$\oplus$317-360 & 45-90$\oplus$270-316\\
$M_{TM_{2}}^{6}$ & IO & 0.003-0.009 & 0.014-0.05 & 0.101-0.111 & 0.048-0.051 & 39.8-51.4 & 0-151$\oplus$211-360 & 19-90$\oplus$270-340\\
 \hline
 \end{tabular}
 \end{small}
\label{table:tm3}
\end{center}
\end{table}

The main results for the neutrino mass matrices with one texture equality and $TM_{2}$ mixing are listed in the following:
\begin{itemize}
\item[i)] Textures $M_{TM_{2}}^{7}, M_{TM_{2}}^{8}$ and $M_{TM_{2}}^{9}$ lead to two degenerate eigenvalues and are, hence, experimentally ruled out at $3\sigma$ C.L.
\item[ii)] Textures $M_{TM_{2}}^{10}, M_{TM_{2}}^{11}, M_{TM_{2}}^{12}, M_{TM_{2}}^{13}, M_{TM_{2}}^{14}$ and $M_{TM_{2}}^{15}$ lead to vanishing reactor mixing angle and, hence, not viable at $3\sigma$ C.L.
\item[iii)] Textures $M_{TM_{2}}^{3}$ and $M_{TM_{2}}^{5}$ for IO are not consistent with the experimental data if cosmological and neutrinoless double beta decay bounds along with neutrino oscillation data are incorporated. 
\item[iv)] Textures $M_{TM_{2}}^{1}$ and $M_{TM_{2}}^{2}$ are consistent with NO only whereas textures $M_{TM_{2}}^{4}$ and $M_{TM_{2}}^{6}$ are consistent with IO only.
\item[v)] For NO, textures $M_{TM_{2}}^{4}$ and $M_{TM_{2}}^{6}$ are not consistent with the experimental data as the mixing angles $\theta_{12}$ and $\theta_{13}$ are not within the $3\sigma$ range.
\item[vi)] All viable textures cannot have zero lowest mass eigenvalue for both mass orderings.
\item[vii)] The atmospheric mixing angle $\theta_{23}$ is below (above) maximal for textures $M_{TM_{2}}^{3}$ ($M_{TM_{2}}^{5}$) and $M_{TM_{2}}^{5}$ ($M_{TM_{2}}^{3}$) with NO and IO, respectively.
\item[viii)] $\theta_{23}$ is maximal for $\delta\sim\frac{\pi}{2}$ or $\frac{3 \pi}{2}$ for textures $M_{TM_{2}}^{1}$ ($M_{TM_{2}}^{6}$) and $M_{TM_{2}}^{2}$($M_{TM_{2}}^{4}$) with NO (IO).   
\item[ix)] The parameter $|M_{ee}|$ is found to be nonzero for all viable textures except $M_{TM_{2}}^{1}$ and $M_{TM_{2}}^{2}$. $|M_{ee}|$ get its largest value when $\delta\sim\frac{\pi}{2}$ or $\frac{3\pi}{2}$ for textures $M_{TM_{2}}^{3}$ and $M_{TM_{2}}^{5}$.
\item[x)] For all viable textures, the effective neutrino mass ($m_{\beta}$) is well within the range provided by KATRIN experiment \cite{katrin}
\item[xi)] The parameters $m_{1}~(m_{3})$, $|M_{ee}|$ and $\sum$ get their largest value when $\theta_{23}\sim45^{\circ}$ for textures $M_{TM_{2}}^{3}$ and $M_{TM_{2}}^{5}$ with NO (IO).

\end{itemize}
\begin{figure}[h]
\begin{center}
\epsfig{file=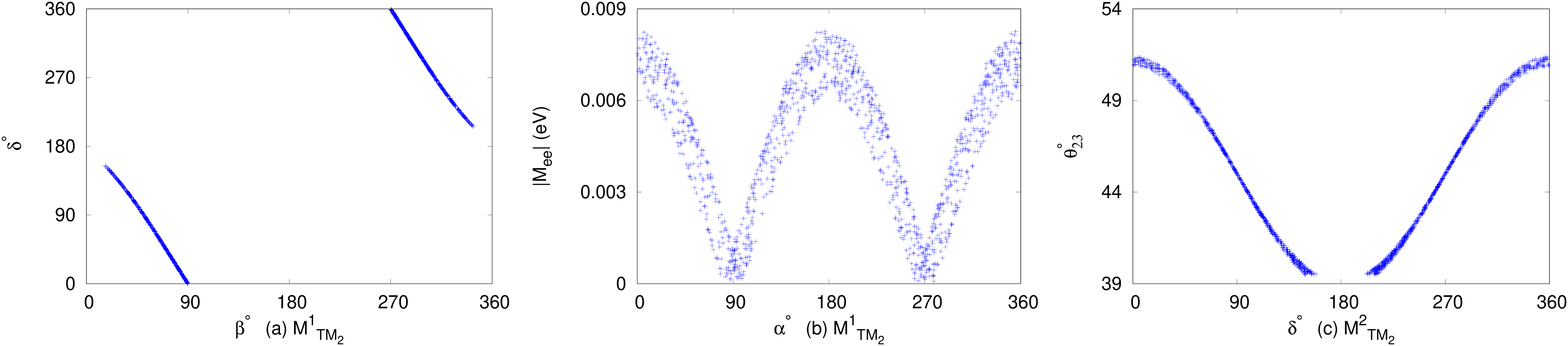, width=15.2cm, height=5.1cm}\\
\epsfig{file=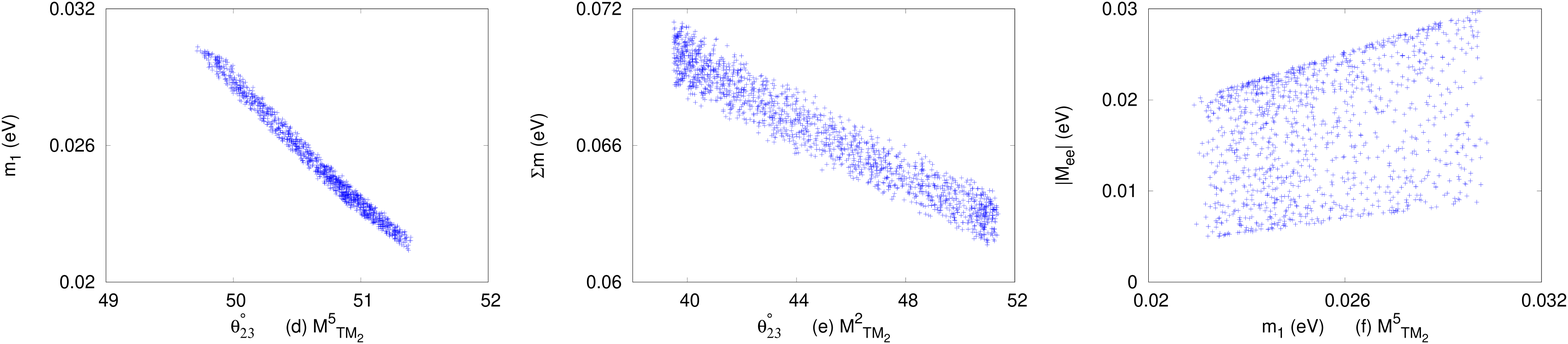, width=15.2cm, height=5.1cm}\\
\epsfig{file=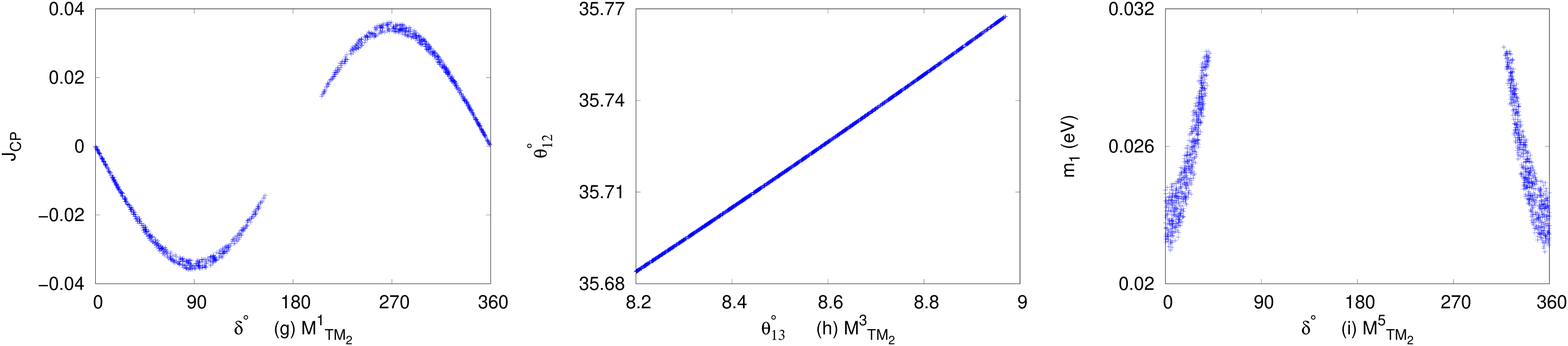, width=15.2cm, height=5.1cm}
\end{center}
\caption{Correlation plots among various parameters for textures (a) $M_{TM_{2}}^{1}$, (b) $M_{TM_{2}}^{1}$, (c) $M_{TM_{2}}^{2}$, (d) $M_{TM_{2}}^{5}$, (e) $M_{TM_{2}}^{2}$ and (f) $M_{TM_{2}}^{5}$, (g) $M_{TM_{2}}^{1}$, (h) $M_{TM_{2}}^{3}$ and (i) $M_{TM_{2}}^{5}$ with NO.}
\end{figure}
\begin{figure}[h]
\begin{center}
\epsfig{file=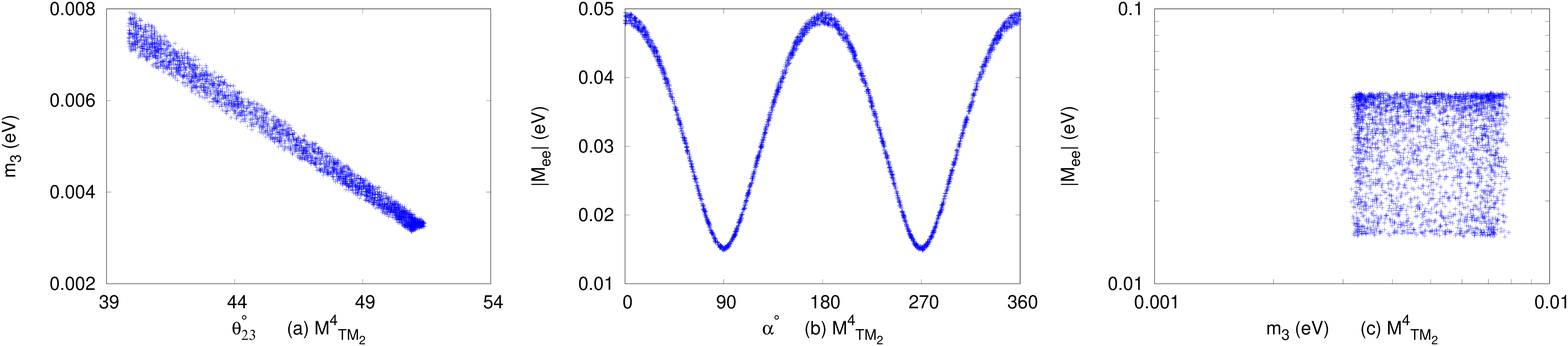, width=15.2cm, height=5.1cm}\\
\epsfig{file=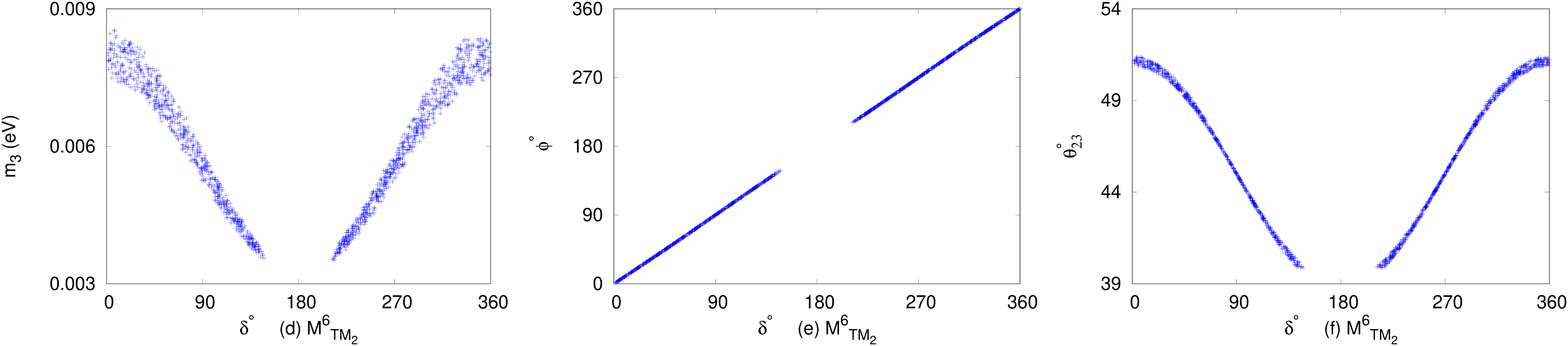, width=15.2cm, height=5.1cm}
\end{center}
\caption{Correlation plots among various parameters for textures (a) $M_{TM_{2}}^{4}$, (b) $M_{TM_{2}}^{4}$, (c) $M_{TM_{2}}^{4}$, (d) $M_{TM_{2}}^{6}$, (e) $M_{TM_{2}}^{6}$ and (f) $M_{TM_{2}}^{6}$ with IO.}
\end{figure}
\subsection{$S_{3}$ Group Motivation}
The $S_{3}$, permutation group of three objects, is the smallest discrete non-Abelian group. The permutation matrices in the three dimensional reducible representation are
\begin{eqnarray}
S^{(1)}&=&\left(
\begin{array}{ccc}
 1 & 0 & 0 \\
 0 & 1 & 0 \\
 0 & 0 & 1 \\
\end{array}
\right),\\
S^{(123)}&=&\left(
\begin{array}{ccc}
 0 & 0 & 1 \\
 1 & 0 & 0 \\
 0 & 1 & 0 \\
\end{array}
\right),~S^{(132)}=\left(
\begin{array}{ccc}
 0 & 1 & 0 \\
 0 & 0 & 1 \\
 1 & 0 & 0 \\
\end{array}
\right),\\
S^{(12)}&=&\left(
\begin{array}{ccc}
 0 & 1 & 0 \\
 1 & 0 & 0 \\
 0 & 0 & 1 \\
\end{array}
\right),~S^{(13)}=\left(
\begin{array}{ccc}
 0 & 0 & 1 \\
 0 & 1 & 0 \\
 1 & 0 & 0 \\
\end{array}
\right),~S^{(23)}=\left(
\begin{array}{ccc}
 1 & 0 & 0 \\
 0 & 0 & 1 \\
 0 & 1 & 0 \\
\end{array}
\right),
\end{eqnarray}
where matrices in each equation belong to the same class of $S_{3}$. The most general neutrino mass matrix invariant under $S_{3}$ group is proportional to the democratic matrix and is given by
\begin{eqnarray}
M_{\nu}=a D  ~~  \textrm{with}~~
D=\left(
\begin{array}{ccc}
 1 & 1 & 1 \\
 1 & 1 & 1 \\
 1 & 1 & 1 \\
\end{array}
\right)
\end{eqnarray}
where $a$ is a complex number and $D$ is called the Democratic matrix. The exact $S_{3}$ symmetry does not satisfy the current neutrino oscillation data and, hence, symmetry must be broken. Various models based on the $S_{3}$ symmetry have been presented in Refs. \cite{s31,s32}. In Ref. \cite{s3}, the $S_{3}$ symmetry is broken by the linear combination of $S_{3}$ group matrices and successfully generates the nonzero $\theta_{13}$.\\
The mass matrices $M_{TM_{2}}$ in Eqs. (21)-(26) can be seen as the linear combination of a democratic part and a symmetry breaking part. The symmetry breaking matrix is the sum of two symmetric matrices out of which one is the $S_{3}$ group matrix which can be any of the $S^{(12)}, S^{(13)}, S^{(23)}$ matrices and the other part is chosen in such a way that the resultant neutrino mass matrix still satisfies the magic symmetry \cite{magic} and remains invariant under $Z_{2}$ symmetry. The mass matrix $M_{TM_{2}}^{1}$ can be rewritten as
\begin{eqnarray}
M_{TM_{2}}^{1}=\left(
\begin{array}{ccc}
 a & a & b \\
 a & b+d & a-d \\
 b & a-d & a+d \\
\end{array}
\right)\equiv a ~D+c ~S^{(13)}+d~ \Delta
\end{eqnarray}
where
\begin{eqnarray}
D=\left(
\begin{array}{ccc}
 1 & 1 & 1 \\
 1 & 1 & 1 \\
 1 & 1 & 1 \\
\end{array}
\right),~S^{(13)}=\left(
\begin{array}{ccc}
 0 & 0 & 1 \\
 0 & 1 & 0 \\
 1 & 0 & 0 \\
\end{array}
\right),~ \Delta=\left(
\begin{array}{ccc}
 0 & 0 & 0 \\
 0 & 1 & -1 \\
 0 & -1 & 1 \\
\end{array}
\right),
\end{eqnarray}
and $a, c, d$ are arbitrary parameters with $b=c+a$.\\
 The $S_{3}$ symmetry of the neutrino mass matrix is broken and the resultant neutrino mass matrix still satisfies $S_{3}$ invariant constraints 
\begin{equation}
M_{\nu_{ii}}-M_{\nu_{jj}}=M_{\nu_{kj}}-M_{\nu_{ki}}~~ \textrm{with}~~i\neq j\neq k. 
\end{equation} 
This leads to a trimaximal eigenvector for the resultant neutrino mass matrix. For example, a typical form of $M_{TM_{2}}^{1}$  neutrino mass matrix is given by
\begin{equation}
M_{TM_{2}}^{1}=\left(
\begin{array}{ccc}
 0.004406\, -0.005578 ~i & 0.004406\, -0.005578~ i & 0.002587\, +0.006938~ i \\
 0.004406\, -0.005578~ i & -0.003415+0.027066~ i & 0.010408\, -0.025705~ i \\
 0.002587\, +0.006938~ i & 0.010408\, -0.025705~ i& -0.001597\, +0.014550 ~i \\
\end{array}
\right).
\end{equation}
In this analysis, we take the charged lepton mass matrix to be diagonal. If a horizontal symmetry exists it must, simultaneously, be a symmetry of the neutrinos as well as the charged leptons before the gauge symmetry breaking. After the symmetry breaking when the fermions acquire nonzero masses, the neutrino sector and the charged lepton sector should be governed by different subgroups of the symmetry group in order to have nonzero mixing. Here, we consider $S_3$ to be the residual symmetry in the neutrino sector and  $Z_{3}$ symmetry as the residual symmetry in the charged lepton sector which yields nondegenerate diagonal charged lepton mass matrix \cite{lam}.\\
Similarly, other viable textures in Eqs. (21)-(23) can, also, be decomposed into the democratic $S_{3}$ invariant part and the symmetry breaking part. The phenomenologically viable mass matrices in Eq. (21)-(23) are related as follows by $S_{3}$ permutation symmetry :
\begin{eqnarray}
&& S^{(123)} M_{TM_{2}}^{1} S^{(123)^T}=M_{TM_{2}}^{4},~~S^{(132)} M_{TM_{2}}^{1} S^{(132)^T}=M_{TM_{2}}^{5},~~S^{(12)} M_{TM_{2}}^{1} S^{(12)^T}=M_{TM_{2}}^{3},\nonumber \\
&& S^{(13)} M_{TM_{2}}^{1} S^{(13)^T}=M_{TM_{2}}^{6},~~S^{(23)} M_{TM_{2}}^{1} S^{(23)^T}=M_{TM_{2}}^{2}.
\end{eqnarray} 
\section{$TM_{1}$ mixing and one texture equality}
The neutrino mixing matrix with first column identical to TBM can be parametrized \cite{tm1} as
\begin{equation}
U_{TM_{1}}=\left(
\begin{array}{ccc}
 \sqrt{\frac{2}{3}} & \frac{\cos\theta}{\sqrt{3}} & \frac{\sin\theta}{\sqrt{3}} \\
 -\frac{1}{\sqrt{6}} & \frac{\cos\theta}{\sqrt{3}}+\frac{e^{i \phi } \sin\theta}{\sqrt{2}} & \frac{\sin\theta}{\sqrt{3}}-\frac{e^{i \phi } \cos\theta}{\sqrt{2}} \\
 -\frac{1}{\sqrt{6}} & \frac{\cos\theta}{\sqrt{3}}-\frac{e^{i \phi } \sin\theta}{\sqrt{2}} & \frac{e^{i \phi } \cos \theta}{\sqrt{2}}+\frac{\sin\theta}{\sqrt{3}} \\
\end{array}
\right)
\end{equation}
and the corresponding neutrino mass matrix is given by
\begin{equation}
M_{TM_{1}}=P_{l}U_{TM_{1}} P_{\nu}M_{\nu}^{diag}P_{\nu}^{T}U_{TM_{1}}^{T}P_{l}^{T}.
\end{equation}
The most general neutrino mass matrix with $TM_{1}$ mixing can be written as
\begin{equation}
M_{TM_{1}}=\left(
\begin{array}{ccc}
 a & 2 b & 2 c \\
 2 b & 4 b+d & a-b-c-d \\
 2 c & a-b-c-d & 4 c+d \\
\end{array}
\right).
\end{equation}
The mass matrix $M_{TM_{1}}$ is invariant under the transformation $G_{1}^{T}M_{TM_{1}}G_{1}=M_{TM_{1}}$
 where $G_{1}=U_{TM_1}diag(1,-1,-1)U^\dagger_{TM_1}$ is the generator of $Z_{2}$ symmetry. This along with equality condition restricts the three unphysical phase angles to $\phi_{e}=\phi_{\mu}=\phi_{\tau}\equiv\phi_{l}$.\\
All possible textures of neutrino mass matrices with $TM_{1}$ mixing and one texture equality are given by 
\begin{eqnarray}
M_{TM_{1}}^{1}&=&\left(
\begin{array}{ccc}
 2 b & 2 b & 2 c \\
 2 b & 4 b+d & b-c-d \\
 2 c & b-c-d & 4 c+d \\
\end{array}
\right),~~~~~~~~~~~~~~~~
M_{TM_{1}}^{2}=\left(
\begin{array}{ccc}
 2 c & 2 b & 2 c \\
 2 b & 4 b+d & -b+c-d \\
 2 c & -b+c-d & 4 c+d \\
\end{array}
\right),\\
M_{TM_{1}}^{3}&=&\left(
\begin{array}{ccc}
 a & 2 b & 2 c \\
 2 b & 2 b & a+b-c \\
 2 c & a+b-c & 4 c-2 b \\
\end{array}
\right),~~~~~~~~~~~~~~~M_{TM_{1}}^{5}=\left(
\begin{array}{ccc}
 a & 2 b & 2 c \\
 2 b & 4 b-2 c & a-b+c \\
 2 c & a-b+c & 2 c \\
\end{array}
\right),\\
M_{TM_{1}}^{4}&=&\left(
\begin{array}{ccc}
 a & 2 b & 2 c \\
 2 b & \frac{1}{2} (a+3 b-c) & \frac{1}{2} (a+3 b-c) \\
 2 c & \frac{1}{2} (a+3 b-c) & \frac{1}{2} (a-5 b+7 c) \\
\end{array}
\right),~~
M_{TM_{1}}^{6}=\left(
\begin{array}{ccc}
 a & 2 b & 2 c \\
 2 b & \frac{1}{2} (a+7 b-5 c) & \frac{1}{2} (a-b+3 c) \\
 2 c & \frac{1}{2} (a-b+3 c) & \frac{1}{2} (a-b+3 c) \\
\end{array}
\right),\\
M_{TM_{1}}^{7}&=&\left(
\begin{array}{ccc}
 a & 2 b & 2 c \\
 2 b & 3 b-c & a \\
 2 c & a & 3 c-b \\
\end{array}
\right),\\
M_{TM_{1}}^{8}&=&\left(
\begin{array}{ccc}
 a & 2 b & 2 c \\
 2 b & 2 c & a+3 b-3 c \\
 2 c & a+3 b-3 c & 6 c-4 b \\
\end{array}
\right),~~~~~~~~~~
M_{TM_{1}}^{9}=\left(
\begin{array}{ccc}
 a & 2 b & 2 c \\
 2 b & 6 b-4 c & a-3 b+3 c \\
 2 c & a-3 b+3 c & 2 b \\
\end{array}
\right),\\
M_{TM_{1}}^{10}&=&\left(
\begin{array}{ccc}
 a & 2 b & 2 c \\
 2 b & a & 3 b-c \\
 2 c & 3 b-c & a-4 b+4 c \\
\end{array}
\right),~~~~~~~~~~~~~~~~~
M_{TM_{1}}^{11}=\left(
\begin{array}{ccc}
 a & 2 b & 2 c \\
 2 b & a+4 b-4 c & 3 c-b \\
 2 c & 3 c-b & a \\
\end{array}
\right),\\
M_{TM_{1}}^{12}&=&M_{TM_{1}}^{13}=\left(
\begin{array}{ccc}
 a & 2 b & 2 b \\
 2 b & 4 b+d & a-2 b-d \\
 2 b & a-2 b-d & 4 b+d \\
\end{array}
\right),\\
M_{TM_{1}}^{14}&=&\left(
\begin{array}{ccc}
 a & 2 b & 2 c \\
 2 b & a+b-c & 2 b \\
 2 c & 2 b & a-3 b+3 c \\
\end{array}
\right),~~~~~~~~~~~~~~
M_{TM_{1}}^{15}=\left(
\begin{array}{ccc}
 a & 2 b & 2 c \\
 2 b & a+3 b-3 c & 2 c \\
 2 c & 2 c & a-b+c \\
\end{array}
\right),
\end{eqnarray}
where textures represented in each equation are related by $\mu$-$\tau$ permutation symmetry. The neutrino mixing angles for $TM_{1}$ mixing in terms of parameters $\theta$ and $\phi$ are \cite{tmzero} given by
\begin{eqnarray}
\sin^{2}\theta_{13}=\frac{1}{3}\sin^{2}\theta,~~ \sin^{2}\theta_{12}=1-\frac{2}{3-\sin^{2}\theta},\\ \nonumber
\textrm{and}~~ \sin^{2}\theta_{23}=\frac{1}{2}\left(1+\frac{\sqrt{6}\sin2\theta \cos\phi}{3-\sin^{2}\theta}\right) .
\end{eqnarray}
For $TM_{1}$ mixing, the Jarlskog rephasing invariant \cite{tmzero} is
\begin{equation}
J_{CP}=\frac{1}{6\sqrt{6}}\sin2\theta \sin\phi,
\end{equation} 
and the Dirac-type CP-violating phase \cite{tmzero} is given by
\begin{equation}
\tan\delta=\frac{\cos2\theta+5}{5\cos2\theta+1}\tan\phi.
\end{equation}
The effective Majorana mass for $TM_{1}$ mixing can be calculated by using Eqs. (12) and (51) as
\begin{equation}
|M_{ee}|=\frac{1}{3}|2m_{1} + m_{2} \cos^{2}\theta e^{2i\alpha} + m_{3} \sin^{2}\theta e^{2i\beta}|,
\end{equation} 
and the effective neutrino mass for $TM_{1}$ by using Eqs.(13) and (51) is given by
\begin{equation}
m_{\beta} = \sqrt{\frac{1}{3}|2m_{1}^2 + m_{2}^2 \cos^{2}\theta + m_{3}^2 \sin^{2}\theta}
\end{equation}

The existence of one equality between the elements of the neutrino mass matrix implies
\begin{equation}
M_{TM_{1}(ab)}-M_{TM_{1}(cd)}=0
\end{equation}  
which yields the following complex equation:
\begin{equation}
\sum(Q V_{ai}V_{bi}-V_{ci}V_{di})~ m_i=0
\end{equation}
where $Q=e^{i(\phi_a+\phi_b-(\phi_c+\phi_d))}$. The above equation can be rewritten as
\begin{equation}
m_1 A_1+m_2 A_2 e^{2i\alpha}+m_3 A_3 e^{2i\beta}=0
\end{equation}
where
\begin{equation}
A_i=(Q U_{ai}U_{bi}-U_{ci} U_{di})
\end{equation}
with $(i=1,2,3)$ and $a,b$ can take values $e, \mu$ and $\tau$. Solving the real and imaginary parts of Eq.(69) simultaneously, we obtain the following two mass ratios:
\begin{small}
\begin{eqnarray}
\zeta\equiv\frac{m_2}{m_1}&=&\frac{(Re(A_1)
  Im(A_3)-Re(A_3) Im(A_1))\cos2\beta+(Re(A_1) Re(A_3)+Im(A_1) Im(A_3))\sin2\beta}{(Re(A_3) Im(A_2)-Re(A_2) Im(A_3))\cos2(\alpha -\beta)+(Re(A_2) Re(A_3)+Im(A_2) Im(A_3))\sin2(\alpha -\beta )},\\
\xi\equiv\frac{m_3}{m_1}&=&\frac{(Re(A_2)
  Im(A_1)-Re(A_1) Im(A_2))\cos2\alpha-(Re(A_1) Re(A_2)+Im(A_1) Im(A_2))\sin2\alpha}{(Re(A_3) Im(A_2)-Re(A_2) Im(A_3))\cos2(\alpha -\beta)+(Re(A_2) Re(A_3)+Im(A_2) Im(A_3))\sin2(\alpha -\beta )}.
\end{eqnarray}
\end{small}
These mass ratios can be used to calculate the ratio of mass squared differences ($R_{\nu}$) which is given by
\begin{equation}
R_{\nu}\equiv\frac{\Delta m^2_{21}}{\Delta m^2_{31}}=\frac{\zeta^2-1}{\xi^2-1}~~\textrm{and}~~ R_{\nu}\equiv\frac{\Delta m^2_{21}}{\Delta m^2_{23}}=\frac{\zeta^2-1}{\zeta^2-\xi^2} 
\end{equation}
for NO and IO, respectively. Since, $\Delta m^2_{21}$ and $\Delta m^2_{31}(\Delta m^2_{23})$ for NO(IO) are experimentally known, the parameter $R_{\nu}$ should lie within its experimentally allowed range for a texture equality to be compatible with the current neutrino oscillation data. The neutrino mass eigenvalues can be calculated by using the relations
\begin{eqnarray}
&& m_2=\sqrt{m_1^2+\Delta m^2_{21}},~~~m_3=\sqrt{m_1^2+\Delta m^2_{31}},\nonumber\\
\textrm{and} && m_2=\sqrt{m_1^2+\Delta m^2_{21}},~~~m_3=\sqrt{m_1^2+\Delta m^2_{21}-\Delta m^2_{23}}
\end{eqnarray}
for NO and IO, respectively.\\
For numerical analysis, we follow the same procedure as  in $TM_{2}$ mixing except that the parameters $\beta$ and $m_{1}$ are generated randomly within their allowed ranges. The mass eigenvalues are calculated by using Eq. (74) and texture equality is imposed by requiring the parameter $R_{\nu}$ in Eq. (73) to lie within its $3\sigma$ experimental range.\\ 
The numerical predictions for unknown parameters are summarized in Table \ref{table:tm11} (where constrains only from neutrino oscillation data are used) and Table \ref{table:tm22} (where the constrains from cosmological and neutrinoless double beta decay bounds along with neutrino oscillation data are used).  The allowed ranges of the parameter 
$\theta_{12}$ are ($34.24^\circ$-$34.42^\circ$) for all viable textures. The parameter $\theta$ is constrained to lie in the ranges ($14.3^\circ$-$15.7^\circ$) whereas $J_{CP}$ lies in the ranges $\pm(0.026$-$0.036)$ for all viable textures. The Majorana phase $\alpha$ varies in the range ($0$-$360^\circ$) for all viable textures with NO only. Correlation plots among various neutrino oscillating parameters are shown in Fig. 3 and Fig. 4 for NO and IO, respectively. $|M_{ee}|$ strongly depends on the Majorana phase $\alpha$ as shown in Fig. 3(a) for NO and Fig. 4(b) for IO. As shown in Fig. 3(d), $\theta_{12}$ is inversely proportional to $\theta_{13}$ which is the classical prediction of $TM_1$ mixing. The Dirac-type CP-violating phase is constrained to lie in the regions around $90^\circ$ and $270^\circ$ which is consistent with the recent observations in the long-baseline neutrino oscillation experiments such as T2K and NOvA \cite{nova} which shows a preference for the Dirac-type CP-violating phase $\delta$ to lie around $\delta\sim 270^\circ$. The main implications for textures having $TM_{1}$ mixing with one texture equality are summarized in the following:
\begin{itemize}
\item[i)] For IO, textures $M_{TM_{1}}^{1}$, $M_{TM_{1}}^{2}$ and $M_{TM_{1}}^{7}$ are not consistent with the neutrino oscillation data at $3\sigma$ C.L.
\item[ii)]For NO, textures $M_{TM_{1}}^{4}$ and $M_{TM_{1}}^{6}$ predict large $\theta_{13}$ and small $\theta_{12}$ and are, hence, experimentally ruled out at $3\sigma$ C.L.
\item[iii)] Textures $M_{TM_{1}}^{12}$ and $M_{TM_{1}}^{13}$ predict a vanishing reactor mixing angle and degenerate mass eigenvalues and are, hence, not viable for both mass orderings.
\item[iv] Textures $M_{TM_{1}}^{3}$, $M_{TM_{1}}^{5}$, $M_{TM_{1}}^{10}$, $M_{TM_{1}}^{14}$, $M_{TM_{1}}^{15}$, $M_{TM_{1}}^{11}$ for both mass orderings and textures $M_{TM_{1}}^{7}$, $M_{TM_{1}}^{8}$ for NO predict large $\sum m_{i}$, and are, hence, not viable with experimental data when cosmological and neutrinoless double beta decay bounds along with neutrino oscillation data are incorporated.
\item[v)] All viable textures cannot have zero lowest mass eigenvalue for both mass orderings.
\item[vi)] The atmospheric mixing angle $\theta_{23}$ is maximal for $\delta\sim\frac{\pi}{2}$ or $\frac{3\pi}{2}$ for all viable textures.
\item[vii)] The parameter $|M_{ee}|$ is found to be bounded from below for all viable textures except $M_{TM_{1}}^{1}$ and $M_{TM_{1}}^{2}$ with NO.
\item[viii)] The parameter $m_{\beta}$ is found to lie within the current experimental range for all viable textures.
\item[ix)] The Dirac-type  CP-violating phase $\delta$ is directly proportional to the parameter $\phi$ for all viable textures. 
\end{itemize}
\begin{scriptsize}
\begin{table}[H]
\begin{center}
\caption{Numerical predictions for viable textures having one equality in $M_{\nu}$ with $TM_{1}$ mixing at $3\sigma$ C.L. (only neutrino oscillation data are incorporated)}
\begin{small}
\begin{tabular}{llllllll}
 \hline\hline
Texture & Ordering & $m_{\textrm{lowest}}$ (eV)& $|M_{ee}|$ (eV)&$\sum$ (eV) & $m_{\beta}$ &$\delta^\circ$& $\beta^\circ$\\
 \hline
$M_{TM_{1}}^{1}$&NO & 0.0039-0.0066 & 0.0-0.0072 & 0.062-0.069 & 0.009-0.012 & 63-125$\oplus$235-297 & 21-65$\oplus$116-159\\ & & & & & & &$\oplus$202-244$\oplus$296-339\\
$M_{TM_{1}}^{2}$ &NO & 0.0039-0.0066 & 0.0-0.0072 & 0.062-0.069 & 0.009-0.012 & 64-125$\oplus$235-296  & 26-68$\oplus$112-154\\ & & & & & & &$\oplus$206-249$\oplus$291-334\\
$M_{TM_{1}}^{3}$ & NO & 0.046-0.67 & 0.027-0.55 & 0.16-2.0 & 0.002-0.003 & 64-122$\oplus$235-297 & 0-43$\oplus$137-222\\ &  & & & & & &$\oplus$318-360\\
&IO & 0.05-0.42 & 0.06-0.41 &  0.21-1.25 &0.077-0.42 &  67-125$\oplus$235-294 & 0-44$\oplus$136-227\\ & & & & & & &$\oplus$314-360\\
$M_{TM_{1}}^{4}$ & IO & 0.007-0.016 & 0.017-0.051 & 0.107-0.121 & 0.048-0.052 & 65-124$\oplus$235-295 & 0-324 \\
$M_{TM_{1}}^{5}$ & NO & 0.04-0.92 & 0.02-0.88  & 0.15-2.8 & 0.045-0.92 & 64-124$\oplus$235-296 & 
 0-36$\oplus$143-218\\ & & & & & & &$\oplus$322-360\\
& IO & 0.062-0.36 & 0.064-0.29 & 0.22-1.1 & 0.079-0.37& 66-125$\oplus$236-293 & 0-35$\oplus$144-213\\ & & & & & & &$\oplus$324-360\\
$M_{TM_{1}}^{6}$ & IO & 0.008-0.017 & 0.017-0.052 & 0.108-0.123 &0.049-0.524 & 65-125$\oplus$234-294 &  0-324\\
$M_{TM_{1}}^{7}$ & NO & 0.085-0.72 & 0.032-0.7  & 0.26-2.2 & 0.085-0.72 & 63-125$\oplus$236-297 & 0-37$\oplus$143-218\\& & & & & & &$\oplus$323-360\\
$M_{TM_{1}}^{8}$ & NO & 0.03-0.32 & 0.02-0.31  & 0.118-0.98 & 0.031-0.32 & 63-124$\oplus$236-297 &  0-53$\oplus$128-235\\ & & & & & & &$\oplus$308-360\\
& IO & 0.014-0.23 & 0.03-0.22 & 0.116-0.7 & 0.05-0.24 & 66-125$\oplus$235-294 & 0-65$\oplus$115-244\\ & & & & & & & $\oplus$296-360\\
$M_{TM_{1}}^{9}$ & NO & 0.025-0.31 & 0.018-0.19  & 0.107-0.93 & 0.025-0.32 & 67-125$\oplus$235-297 &  0-50$\oplus$130-224\\ & & & & & & &$\oplus$312-360\\
& IO & 0.0148-0.33 & 0.028-0.26 & 0.117-1.0 & 0.05-0.33 & 66-125$\oplus$235-294 & 0-56 $\oplus$124-236\\ & & & & & & &$\oplus$306-360\\
$M_{TM_{1}}^{10}$ & NO & 0.034-0.7 & 0.014-0.65  & 0.13-2.1 & 0.034-0.7 & 63-124$\oplus$237-297 & 43-141$\oplus$221-314\\
& IO & 0.028-0.4 & 0.021-0.27 & 0.14-1.2 & 0.05-0.41 & 66-125$\oplus$235-295 & 34-145$\oplus$215-325\\
$M_{TM_{1}}^{11}$ & NO & 0.031-0.3 & 0.011-0.27  & 0.122-0.86 & 0.032-0.29 & 65-125$\oplus$235$-$295 &  47-128$\oplus$229-313\\
& IO & 0.03-0.51 & 0.023-0.48 & 0.14-1.52 & 0.05-0.51 & 65-124$\oplus$236-295 & 44-136$\oplus$224-316\\
$M_{TM_{1}}^{14}$ & NO & 0.04-0.48 & 0.02-0.36  & 0.16-1.43 & 0.04-0.48 & 63-123$\oplus$237-297 &  47-132$\oplus$230-314\\
& IO & 0.06-0.4 & 0.028-0.38 & 0.21-1.2 & 0.07-0.41 & 65-126$\oplus$234-294 & 43-135$\oplus$225-316\\
$M_{TM_{1}}^{15}$ & NO & 0.04-0.55 & 0.01-0.5  & 0.15-1.7 & 0.04-0.55 & 64-123$\oplus$236-295 & 53-125$\oplus$237-307\\
& IO & 0.06-0.52 & 0.03-0.5 & 0.22-1.6 & 0.07-0.52 & 65-125$\oplus$235-295 & 54-126$\oplus$234-305\\
 \hline
 \end{tabular}
 \end{small}
\label{table:tm11}
\end{center}
\end{table}
\end{scriptsize}

\begin{scriptsize}
\begin{table}[H]
\begin{center}
\caption{Numerical predictions for viable textures having one equality in $M_{\nu}$ with $TM_{1}$ mixing at $3\sigma$ C.L. (cosmological and neutrinoless double beta decay bounds along with neutrino oscillation data are incorporated)}
\begin{small}
\begin{tabular}{llllllll}
 \hline\hline
Texture & Ordering & $m_{\textrm{lowest}}$ (eV)& $|M_{ee}|$ (eV)&$\sum$ (eV)& $m_{\beta}$ &$\delta^\circ$ & $\beta^\circ$\\
 \hline
$M_{TM_{1}}^{1}$&NO & 0.0039-0.0066 & 0.0-0.0072 & 0.062-0.069 & 0.009-0.012 & 64-125$\oplus$235-296 & 21-65$\oplus$116-158\\ & & & & & &$\oplus$201-245$\oplus$295-339\\
$M_{TM_{1}}^{2}$ &NO & 0.0039-0.0066 & 0.0-0.0072 & 0.062-0.069 & 0.009-0.012 & 64-125$\oplus$235-296 & 26-68$\oplus$112-154\\ & & & & & &$\oplus$206-249$\oplus$291-334\\
$M_{TM_{1}}^{4}$ & IO & 0.007-0.016 & 0.017-0.051 & 0.107-0.12 & 0.048-0.052 & 66-125$\oplus$235-295 & 0-322 \\
$M_{TM_{1}}^{6}$ & IO & 0.008-0.017 & 0.017-0.051 & 0.108-0.12 & 0.049-0.052 & 65-125$\oplus$235-295 & 0-320\\
$M_{TM_{1}}^{8}$ & IO & 0.014-0.017 & 0.038-0.049 & 0.116-0.12 & 0.05-0.052 & 84-124$\oplus$236-273 & 11-53$\oplus$125-170\\ & & & & & & & $\oplus$193-235$\oplus$307-347\\
$M_{TM_{1}}^{9}$ & NO & 0.025-0.031 & 0.018-0.031  & 0.107-0.12 & 0.026-0.32 & 112-125$\oplus$235-247 &  1-29$\oplus$152-208\\ & & & & & & &$\oplus$332-359\\
& IO & 0.0148-0.0167 & 0.037-0.047 & 0.117-0.12 & 0.05-0.052 & 67-95$\oplus$269-293 & 16-44$\oplus$135-168\\ & & & & & & &$\oplus$191-225$\oplus$316-348\\

 \hline
 \end{tabular}
 \end{small}
\label{table:tm22}
\end{center}
\end{table}
\end{scriptsize}

\begin{figure}[h]
\begin{center}
\epsfig{file=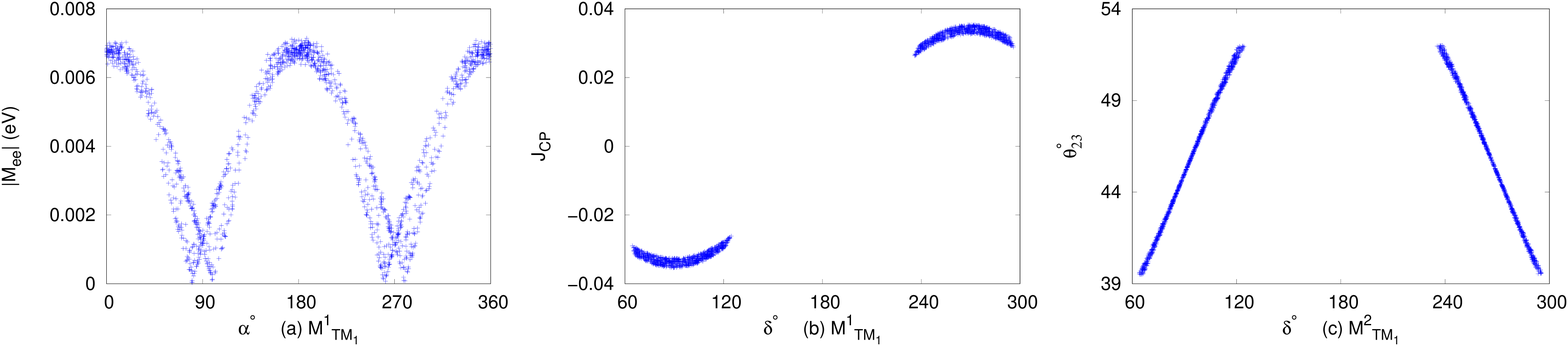, width=15.2cm, height=5.1cm}\\
\epsfig{file=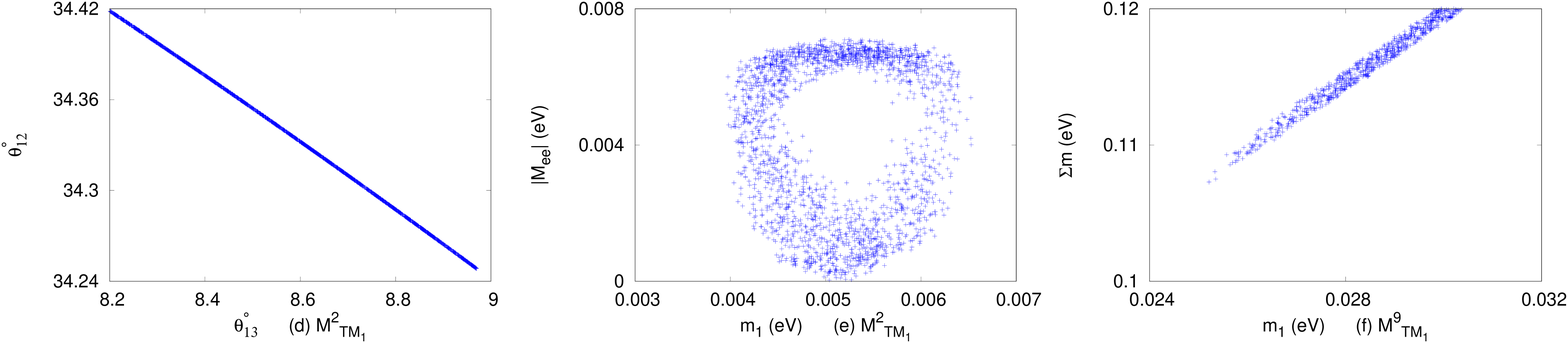, width=15.2cm, height=5.1cm}
\end{center}
\caption{Correlation plots among various parameters for textures (a) $M_{TM_{1}}^{1}$, (b) $M_{TM_{1}}^{1}$, (c) $M_{TM_{1}}^{2}$, (d) $M_{TM_{1}}^{2}$, (e) $M_{TM_{1}}^{2}$ and (f) $M_{TM_{1}}^{9}$ with NO.}
\end{figure}
\begin{figure}[h]
\begin{center}
\epsfig{file=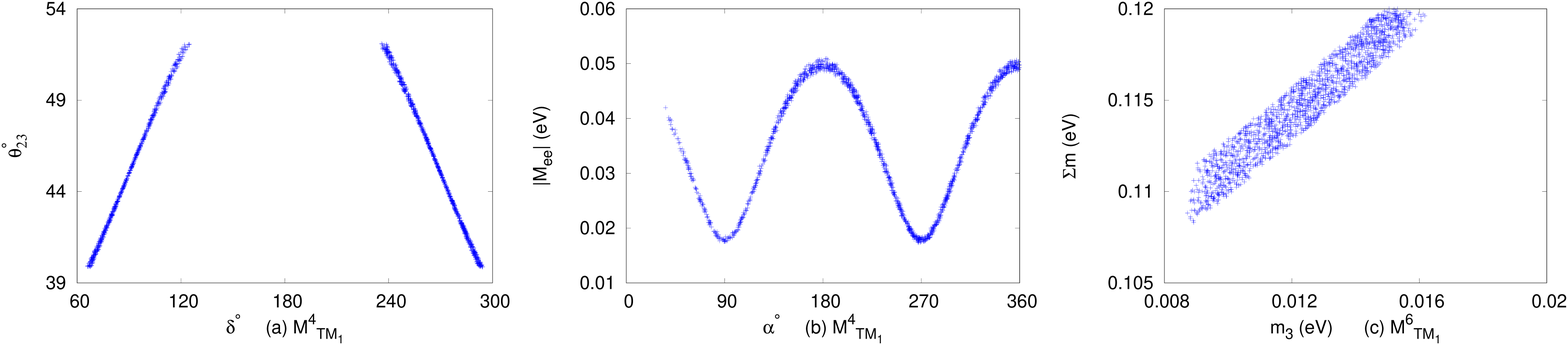, width=15.2cm, height=5.1cm}\\
\epsfig{file=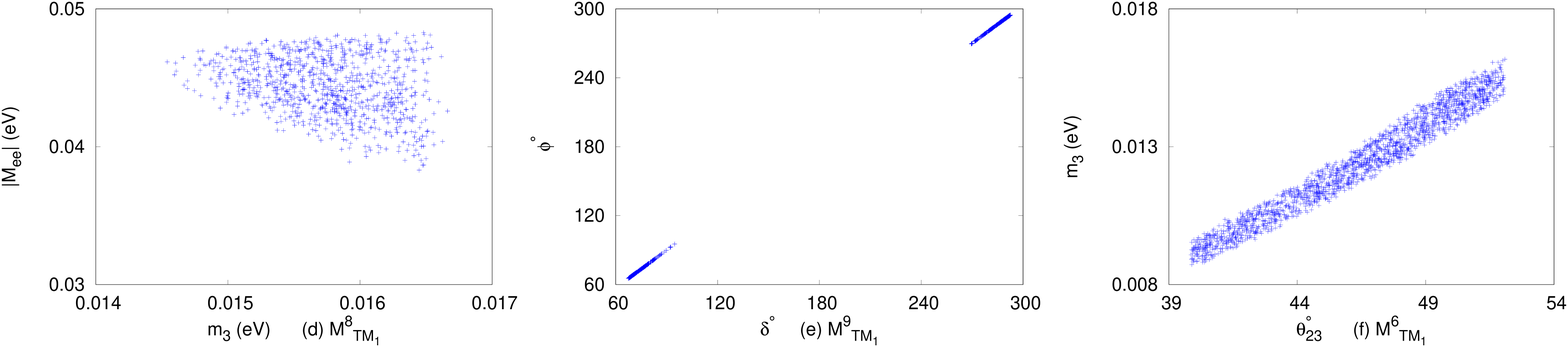, width=15.2cm, height=5.1cm}
\end{center}
\caption{Correlation plots among various parameters for textures (a) $M_{TM_{1}}^{4}$, (b) $M_{TM_{1}}^{4}$, (c) $M_{TM_{1}}^{6}$, (d) $M_{TM_{1}}^{8}$, (e) $M_{TM_{1}}^{9}$ and (f) $M_{TM_{1}}^{6}$ with IO.}
\end{figure}

\section{Summary}
We studied the phenomenological implications of one texture equality in the neutrino mass matrix with $TM_{1}$ or $TM_{2}$ mixing. The presence of one texture equality in $M_{\nu}$ with $TM_{1}$ or $TM_{2}$ as the mixing matrix reduces the number of free parameters significantly and, hence, leads to very predictive neutrino mass matrices. Out of total fifteen possible textures of $M_{\nu}$, thirteen textures are phenomenologically allowed with $TM_{1}$ mixing and only six textures are allowed with $TM_{2}$ mixing in the light of current neutrino oscillation data at $3\sigma$ C.L. However, the number of viable textures reduced to six for $TM_{1}$ mixing if the constrains from cosmology and neutrinoless double beta decay experiments along with neutrino oscillation data are used. Since, the $TM_{2}$ mixing predicts a value of $\theta_{12}$ away from its best fit value, $TM_{1}$ mixing is phenomenologically more appealing. In this analysis, we have obtained interesting predictions for unknown parameters such as the Dirac- and Majorana-type CP-violating phases, effective Majorana neutrino mass, effective neutrino mass, Jarlskog rephasing invariant, neutrino mass scale and the sum of neutrino masses. For $TM_{1}$ mixing, the Dirac-type CP-violating phase $(\delta)$ is restricted to the regions around $\frac{\pi}{2}$ and $\frac{3\pi}{2}$, the atmospheric mixing angle $(\theta_{23})$ is maximal for $\delta\sim\frac{\pi}{2}$ or $\frac{3\pi}{2}$ and the lowest neutrino mass eigenvalue cannot be zero for all viable textures. For $TM_{2}$ mixing, the CP-violating phases $\delta$ and $\beta$ are strongly correlated, $\theta_{23}$ is below (above) maximal for textures $M_{TM_{2}}^{3}(M_{TM_{2}}^{5})$ and $M_{TM_{2}}^{5}(M_{TM_{2}}^{3})$ with NO and IO, respectively and the lowest mass eigenvalue cannot be zero for all viable textures. For $M_{TM_{2}}$ mass matrices with one texture equality, the residual $S_{3}$ symmetry is broken and resulting neutrino mass matrix is invariant under $Z_2$ symmetry.   

\section{Acknowledgements}
The research work of S. D. is supported by the Council of Scientific and Industrial Research (CSIR), Government of India, New Delhi vide grant No. 03(1333)/15/EMR-II. S. D. gratefully acknowledges the kind hospitality provided by IUCAA, Pune. Authors thank Radha Raman Gautam and Lal Singh for carefully reading the manuscript. A previous version of this manuscript was presented in arXiv.org with identifier 2202.13070.

\end{document}